\documentclass{cimento}

\title{The role of pomeron states in the Glueball Resonance Gas}
\author{E.~Trotti\from{ins:x}\thanks{trottienrico@gmail.com} \atque
S.~Jafarzade\from{ins:x}\from{ins:y}\thanks{shahriyar.jzade@gmail.com}}
\instlist{\inst{ins:x} Institute of Physics, Jan Kochanowski University,
ul. Uniwersytecka 7, 25-406, Kielce, Poland\\
\inst{ins:y} Institute of Radiation Problems, Ministry of Science and Education, B.Vahabzade St. 9, AZ1143, Baku, Azerbaijan}
\begin{document}

\maketitle

\begin{abstract}
In this note, we present our recent analyses of the thermodynamic properties of the glueball resonance gas. We observe that the dominant contribution to the thermodynamic quantities, such as pressure, trace anomaly, and entropy, is coming from the free glueball gas with the states of positive charge conjugation (i.e., pomeron). A comparison of pomeron states obtained from LQCD and functional methods within the glueball resonance gas model is also presented.
\end{abstract}

\section{Introduction}
\label{sec.intro}
Quantum chromodynamics (QCD) can be described in the high-temperature regime by a perturbative quark-gluon plasma, while in the low-temperature regime, by a gas of weakly interacting hadrons. The two regimes are separated by the pseudo-critical temperature $T_c$, which represents the confinement-deconfinement cross-over transition of QCD whose value is still under debate  \cite{Panero:2009tv,Lucini:2012gg,Borsanyi:2012ve,Caselle:2011fy,Caselle:2015tza}. In the case of Yang-Mills (YM) (QCD without quarks), the $T_c$ represents the transition (of the first order) between the bound states of gluons (glueballs) and gas of gluons at low and high temperatures, respectively.

At finite temperature, several approaches have been developed to study QCD \cite{Pilaftsis:2013xna,Koenigstein:2021syz,Broniowski:2015oha,Samanta:2021vgt}.
%\st{Among them, in this work primary importance is given to lattice QCD (LQCD) and to the hadron resonance gas model (HRG)}.
One typically compares the outcomes of the lattice QCD (LQCD) to the hadron resonance gas model (HRG) below $T_c$. This procedure can be repeated in the YM case. Instead of HRG, one has a subset of it, that we call a Glueball Resonance Gas (GRG) \cite{Trotti:2022knd}. 

The study on the pomeron and odderon trajectories suggests the existence of glueballs \cite{Szanyi:2019kkn,Godizov:2016vuw}. %From another point of view, according to the recent phenomenological analyses this resonance deserves to be the glueball state among the other resonances \cite{Vereijken:2023jor}.
We observe that,
within the GRG, the contribution provided by the pomeron, $J^{P+}$, states is predominant, since the contribution to the thermodynamic (TD) quantities, e.g., pressure and trace anomaly, is dominated by the lightest resonances ($0^{++} $ and $2^{++}$). 
%\st{This study has already been done in }\cite{Trotti:2022knd}, where the GRG model was applied to three lattice spectra \cite{Chen:2005mg,Meyer:2004gx,Athenodorou:2020ani} and then compared with the lattice thermodynamics data from Ref. \cite{Borsanyi:2012ve}.

We have studied the GRG in \cite{Trotti:2022knd} by considering the mass spectra of glueballs from Refs. \cite{Chen:2005mg,Meyer:2004gx,Athenodorou:2020ani}. %\st{Within this framework, one can also consider other glueball spectra not produced with LQCD methods}.
In this work, we compare the results for the TD quantities coming from the pomeron states obtained from the LQCD  \cite{Athenodorou:2020ani} and the functional method \cite{Huber:2021yfy}.

\section{Results}
\label{sec.Perrors}
The TD of the GRG can be described by using the total, dimensionless energy density $\hat{\epsilon}$ 
\begin{equation}
\hat{\epsilon} =\sum_{i=1}^{N}\dfrac{(2J_i+1)}{T^4}\int_0^{\infty} \dfrac{k^2}{2 \pi^2} \dfrac{\sqrt{k^2 +m_i^2}}{\exp \Biggl[ \dfrac{\sqrt{k^2 +m_i^2}}{T}  \Biggr] - 1} dk \text{ ,}
\end{equation}
and the pressure $\hat{p}$:
\begin{equation}
    \hat{p}=-\sum_{i=1}^{N}\dfrac{(2J_i+1)}{T^3}\int_{0}^\infty \dfrac{k^2}{2 \pi^2} \ln{\Big(1-e^{-\frac{\sqrt{k^2+m_i^2}}{T}}\Big)}dk \text{ ,}
    \label{eq.pfree}
\end{equation}
where $J_i$ is the total spin of the $i$-th state. In the description of the GRG, other two quantities are also relevant, i.e., the dimensionless trace anomaly $\hat{I}$ and the entropy density $\hat{s}$:
\begin{equation}
    \hat{I}=\hat{\epsilon}-3\hat{p}\,,\quad \hat{s} = \hat{p}+\hat{\epsilon} \text{ .}
    \label{IS}
\end{equation}

\begin{figure}[!htb]
\centering
  \includegraphics[scale=0.6]{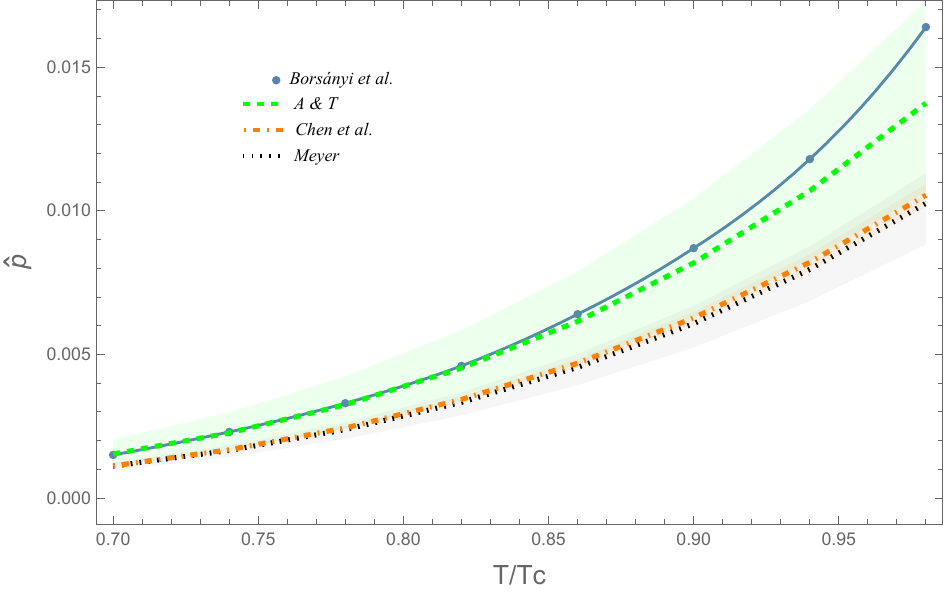}
  \caption{Pressure of the GRG as a function of the temperature for three different sets
of the LQCD mass spectra \cite{Chen:2005mg,Meyer:2004gx,Athenodorou:2020ani},
compared to the same quantity evaluated in Ref. \cite{Borsanyi:2012ve}.}\label{fig:errors}
\end{figure}

%\section{Results}
These quantities play a central role in Ref. \cite{Trotti:2022knd}, where the comparison between 
the LQCD TD data from Ref. \cite{Borsanyi:2012ve} and GRG model constructed the lattice spectra from \cite{Chen:2005mg,Meyer:2004gx,Athenodorou:2020ani} was performed. The results have shown that the GRG with the most recent lattice work on the glueball masses \cite{Athenodorou:2020ani} better describes LQCD TD data. In Fig. \ref{fig:errors}, this comparison is shown for the normalized pressure, with the addition of the statistical errors (not present in Ref. \cite{Trotti:2022knd}).

\noindent
Even when considering the errors in the pressure, we confirm the masses from Ref. \cite{Athenodorou:2020ani} are favored. We remind that the TD results provided in Ref. \cite{Borsanyi:2012ve} are given in functions of  $T/T_c$; thus, the same quantities calculated from the GRG model must be presented for this ratio. However, lattice parameters used to calculate the mass spectra in \cite{Chen:2005mg,Meyer:2004gx,Athenodorou:2020ani} imply different $T_c$ values for each line in Fig \ref{fig:errors}. For more details, see Ref. \cite{Trotti:2022knd}.

\begin{figure}[!htb]
\centering
  \includegraphics[width=8cm]{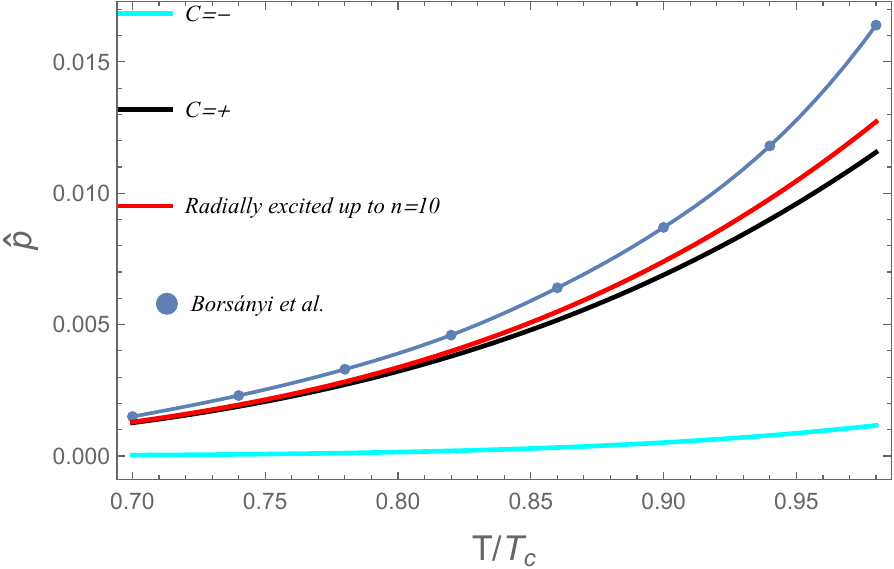}
  \caption{Pressure obtained in Ref. \cite{Borsanyi:2012ve} compared to the GRG by considering the spectrum of Ref. \cite{Athenodorou:2020ani} and all excited stated up to $n=10$ (red), the pomeron ($C=+$) states (black) and the odderon ($C=-$) states (cyan).}\label{fig:pomeron}
\end{figure}

In Ref \cite{Trotti:2022knd}, we included the excited
states by using Regge trajectory fitted from the glueballs with the quantum numbers $J^{PC}=0^{++},2^{++},0^{-+},2^{-+},1^{+-}$ using the spectrum of Ref. \cite{Athenodorou:2020ani}. The results do not significantly change, even with the addition of states up to radial quantum number $n=10$. As we see from Fig. \ref{fig:pomeron}, among these states, the ones with
positive charge conjugation provide the main contribution to the pressure, while those with C=-1 have a negligible effect up to the vicinity of $T_c$. 

%\section{The pomeron effect}
%\label{sec.Pomer}
As an additional task, we compare the (favored) masses of \cite{Athenodorou:2020ani} with the mass predictions obtained with functional methods \cite{Huber:2021yfy}. The comparison is possible by neglecting the odderon contribution from \cite{Athenodorou:2020ani}. In Table \ref{tablemass}, the masses that will be used in the GRG model are reported.

\begin{figure}[!htb]
\centering
\minipage{0.45\textwidth}
  \includegraphics[width=6cm]{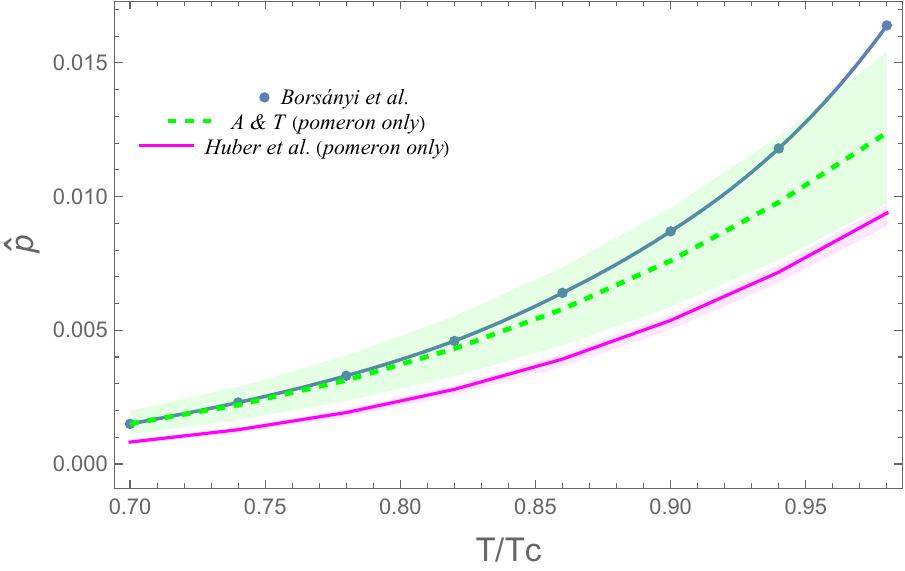}
\endminipage\hfill
\minipage{0.45\textwidth}
  \includegraphics[width=6cm]{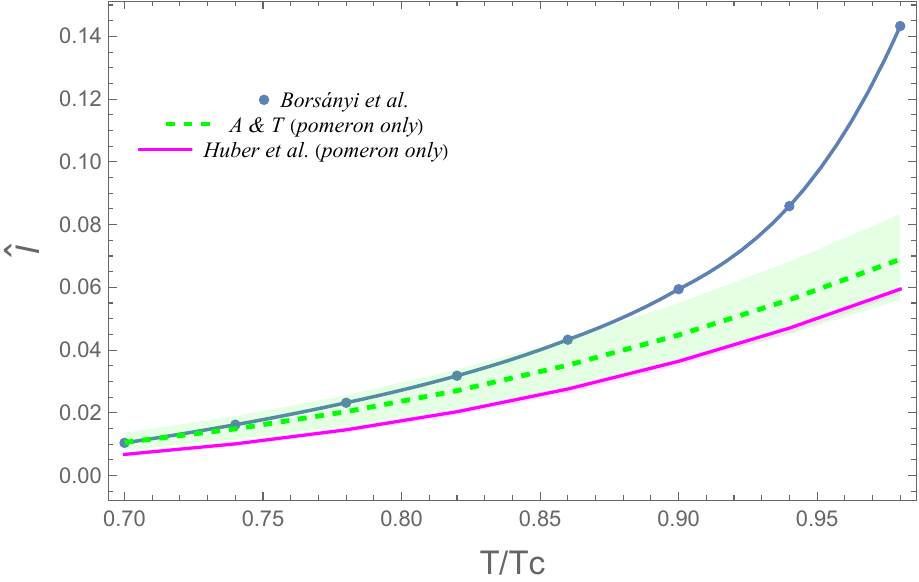}
\endminipage\hfill
\minipage{0.45\textwidth}
%\centering
\includegraphics[width=6cm]{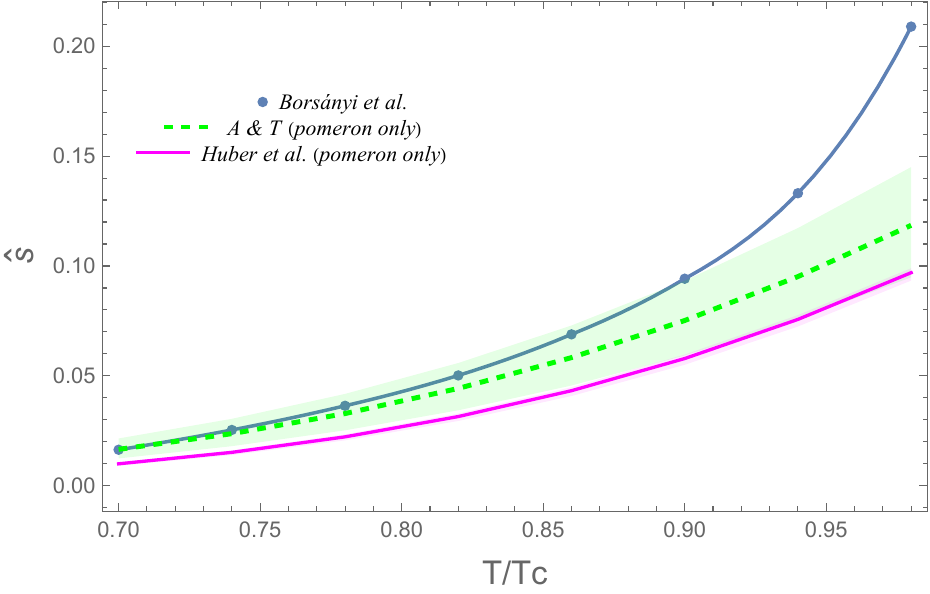}
\endminipage
  \caption{Pressure (top, left), trace anomaly (top, right), and entropy (bottom) as a function of $T/T_c$. The green dashed, and magenta continuous, plots (from the GRG model with the spectra (pomerons only) from \cite{Athenodorou:2020ani} and \cite{Huber:2021yfy} respectively) are reported with errors due to the mass uncertainties. The comparison is done with the lattice data from \cite{Borsanyi:2012ve}.}\label{fig:plots}
\end{figure}

\begin{table}[ht]
\small
 % \begin{center}
  
 \begin{tabular}{|c|c|c|c|c|c|}
    \hline
    \rule{0pt}{2em} $n\,J^{PC}$ & \multicolumn{2}{c|}{M[MeV]} &  \rule{0pt}{2em} $n\,J^{PC}$ & \multicolumn{2}{c|}{M[MeV]}\\ \hline
     & Huber et al. \cite{Huber:2021yfy}  &A \& T \cite{Athenodorou:2020ani}  & & Huber et al. \cite{Huber:2021yfy}  &A \& T \cite{Athenodorou:2020ani}  \\\hline
     $\textbf{1\, 0}^{++}$ & 1850(130)& 1653(26) &   $\textbf{1\,0}^{-+}$ &  2580(180)  & 2561(40)
 \\\hline
     $\textbf{2\,0}^{++}$ & 2570(210)  & 2842(40) &  $\textbf{2\,0}^{-+}$ & 3870(120)   & 3540(80) 
   \\\hline
    $\textbf{1\,2}^{++}$ & 2610(180)& 2376(32) &  $\textbf{1\,2}^{-+}$ & 2740(140)& 3070(60)
 \\\hline
    $\textbf{2\,2}^{++}$ &  3640(240) &  3300(50) &   $\textbf{2\,2}^{-+}$ & 4300(190) & 3970(70)
 \\\hline
    $\textbf{1\,3}^{++}$ & 3370(50) & 3740(70)&  $\textbf{1 4}^{++}$ &  4140(30)   & 3690(80)
 \\\hline
    \end{tabular}  
    \caption{The pomeron spectra reported in Refs. \cite{Athenodorou:2020ani,Huber:2021yfy}.}
     \label{tablemass}
 %\end{center}
 \end{table}
 \noindent
In Refs. \cite{Athenodorou:2020ani,Huber:2021yfy}, the values are reported using the same value of the lattice parameter $r_0^{-1}=418(5)$ MeV. By considering the relation between $T_c$ and $r_0$ \cite{Gockeler:2005rv}:
\begin{equation}
     T_c= 1.26(7) \cdot 0.614(2) \cdot r_0^{-1}  \text{ ,}\label{eq:TC1}
\end{equation}
we obtain the common value $T_c=323\pm 18 $ MeV. Three TD values (pressure, trace anomaly and entropy) are shown in Fig. \ref{fig:plots}. One can see that the values obtained using the GRG model with the masses from \cite{Huber:2021yfy} are lower than those from \cite{Athenodorou:2020ani}. This is due mainly to the effect of the lightest states since a slight increase in the mass of the $i$-th glueball is reflected in a sizeable decrease in the TD quantities. %Still, the large errors reported by Huber et al. allow a small superposition of the values. 
However, in both cases, a slight discrepancy is still present with the lattice TD results \cite{Borsanyi:2012ve} (which include the total, odderon, and pomeron contribution).

%\begin{figure}[!htb]
%\centering
 % \includegraphics[width=8cm]{presATHU.pdf}
  %\caption{Pressure as function of $T/T_c$. The green, dashed, and magenta, continuous, plots (from the GRG model with the spectra from \cite{Athenodorou:2020ani} and \cite{Huber:2021yfy} respectively) are reported with the error bar due to the mass uncertainties. The comparison is done with the lattice data from \cite{Borsanyi:2012ve}.}\label{fig:pomeron}
%\end{figure}

%\section{Effect of the $2^{++}$ interaction}

\section{Conclusions}
We revisited the results reported in Ref. \cite{Trotti:2022knd} by considering the statistical errors in the mass spectrum, confirming that GRG with the glueball spectrum from  \cite{Athenodorou:2020ani} better describes the LQCD date from \cite{Borsanyi:2012ve}. Among the states, the ones with positive charge conjugation (pomeron) provide the dominant contribution. Along this line, we present the comparison between the LQCD spectra of \cite{Athenodorou:2020ani} with the one observed by functional methods \cite{Huber:2021yfy}, and we evaluate TD quantities for both pomeron spectra within GRG. %by considering only the pomeron states obtained in  within GRG. The corresponding plots for the thermodynamic properties are depicted.  

%This is also supported by the new plots where the error bars are taken into account. Since the pomeron part gives the largest contribution to the thermodynamic quantities, we compared these quantities obtained with the GRG from two glueball spectra (lattice and functional methods). The masses from \cite{Athenodorou:2020ani}, chosen to represent the lattice spectra, confirmed their attainability also when compared with the masses from \cite{Huber:2021yfy}.

\newpage

\acknowledgments
The authors thank F. Giacosa for previous publications \cite{Trotti:2022knd,Giacosa:2021brl} and for useful comments. 
The authors acknowledge financial support through the project AKCELERATOR ROZWOJU
Uniwersytetu Jana Kochanowskiego w Kielcach (Development Accelerator of the
Jan Kochanowski University of Kielce), co-financed by the European Union
under the European Social Fund, with no. POWR.03.05.00-00-Z212/18. The work of Shahriyar Jafarzade is partially supported by the Polish National Science Centre (NCN) through the OPUS
project 2019/33/B/ST2/00613.


\begin{thebibliography}{0}

%\cite{Panero:2009tv}
\bibitem{Panero:2009tv}
M.~Panero,
%``Thermodynamics of the QCD plasma and the large-N limit,''
Phys. Rev. Lett. \textbf{103} (2009), 232001
doi:10.1103/PhysRevLett.103.232001
[arXiv:0907.3719 [hep-lat]].
%326 citations counted in INSPIRE as of 09 Oct 2023

%\cite{Lucini:2012gg}
\bibitem{Lucini:2012gg}
B.~Lucini and M.~Panero,
%``SU(N) gauge theories at large N,''
Phys. Rept. \textbf{526} (2013), 93-163
doi:10.1016/j.physrep.2013.01.001
[arXiv:1210.4997 [hep-th]].
%252 citations counted in INSPIRE as of 09 Oct 2023

%\cite{Borsanyi:2012ve}
\bibitem{Borsanyi:2012ve}
S.~Borsanyi, G.~Endrodi, Z.~Fodor, S.~D.~Katz and K.~K.~Szabo,
%``Precision SU(3) lattice thermodynamics for a large temperature range,''
JHEP \textbf{07} (2012), 056
doi:10.1007/JHEP07(2012)056
[arXiv:1204.6184 [hep-lat]].
%248 citations counted in INSPIRE as of 09 Oct 2023
%\cite{Caselle:2011fy}
\bibitem{Caselle:2011fy}
M.~Caselle, L.~Castagnini, A.~Feo, F.~Gliozzi and M.~Panero,
%``Thermodynamics of SU(N) Yang-Mills theories in 2+1 dimensions I - The confining phase,''
JHEP \textbf{06} (2011), 142
doi:10.1007/JHEP06(2011)142
[arXiv:1105.0359 [hep-lat]].
%43 citations counted in INSPIRE as of 09 Oct 2023
%\cite{Caselle:2015tza}
\bibitem{Caselle:2015tza}
M.~Caselle, A.~Nada and M.~Panero,
%``Hagedorn spectrum and thermodynamics of SU(2) and SU(3) Yang-Mills theories,''
JHEP \textbf{07} (2015), 143
[erratum: JHEP \textbf{11} (2017), 016]
doi:10.1007/JHEP07(2015)143
[arXiv:1505.01106 [hep-lat]].
%42 citations counted in INSPIRE as of 09 Oct 2023

%\cite{Pilaftsis:2013xna}
\bibitem{Pilaftsis:2013xna}
A.~Pilaftsis and D.~Teresi,
%``Symmetry Improved CJT Effective Action,''
Nucl. Phys. B \textbf{874} (2013) no.2, 594-619
doi:10.1016/j.nuclphysb.2013.06.004
[arXiv:1305.3221 [hep-ph]].
%49 citations counted in INSPIRE as of 09 Oct 2023
%\cite{Koenigstein:2021syz}
\bibitem{Koenigstein:2021syz}
A.~Koenigstein, M.~J.~Steil, N.~Wink, E.~Grossi, J.~Braun, M.~Buballa and D.~H.~Rischke,
%``Numerical fluid dynamics for FRG flow equations: Zero-dimensional QFTs as numerical test cases. I. The O(N) model,''
Phys. Rev. D \textbf{106} (2022) no.6, 065012
doi:10.1103/PhysRevD.106.065012
[arXiv:2108.02504 [cond-mat.stat-mech]].
%26 citations counted in INSPIRE as of 09 Oct 2023
%\cite{Broniowski:2015oha}
\bibitem{Broniowski:2015oha}
W.~Broniowski, F.~Giacosa and V.~Begun,
%``Cancellation of the $\sigma$ meson in thermal models,''
Phys. Rev. C \textbf{92} (2015) no.3, 034905
doi:10.1103/PhysRevC.92.034905
[arXiv:1506.01260 [nucl-th]].
%74 citations counted in INSPIRE as of 09 Oct 2023
%\cite{Samanta:2021vgt}
\bibitem{Samanta:2021vgt}
S.~Samanta and F.~Giacosa,
%``Role of bound states and resonances in scalar QFT at nonzero temperature,''
Phys. Rev. D \textbf{107} (2023) no.3, 036001
doi:10.1103/PhysRevD.107.036001
[arXiv:2110.14752 [hep-ph]].

%\cite{Trotti:2022knd}
\bibitem{Trotti:2022knd}
E.~Trotti, S.~Jafarzade and F.~Giacosa,
%``Thermodynamics of the glueball resonance gas,''
Eur. Phys. J. C \textbf{83} (2023) no.5, 390
doi:10.1140/epjc/s10052-023-11557-0
[arXiv:2212.03272 [hep-ph]].
%1 citations counted in INSPIRE as of 09 Oct 2023

%\cite{Szanyi:2019kkn}
\bibitem{Szanyi:2019kkn}
I.~Szanyi, L.~Jenkovszky, R.~Schicker and V.~Svintozelskyi,
%``Pomeron/glueball and odderon/oddball trajectories,''
Nucl. Phys. A \textbf{998} (2020), 121728
doi:10.1016/j.nuclphysa.2020.121728
[arXiv:1910.02494 [hep-ph]].

%\cite{Godizov:2016vuw}
\bibitem{Godizov:2016vuw}
A.~A.~Godizov,
%``The ground state of the Pomeron and its decays to light mesons and photons,''
Eur. Phys. J. C \textbf{76} (2016) no.7, 361
doi:10.1140/epjc/s10052-016-4229-z
[arXiv:1604.01689 [hep-ph]].

%\cite{Chen:2005mg}
\bibitem{Chen:2005mg}
Y.~Chen, A.~Alexandru, S.~J.~Dong, T.~Draper, I.~Horvath, F.~X.~Lee, K.~F.~Liu, N.~Mathur, C.~Morningstar and M.~Peardon, \textit{et al.}
%``Glueball spectrum and matrix elements on anisotropic lattices,''
Phys. Rev. D \textbf{73} (2006), 014516
doi:10.1103/PhysRevD.73.014516
[arXiv:hep-lat/0510074 [hep-lat]].
%638 citations counted in INSPIRE as of 09 Oct 2023
%\cite{Meyer:2004gx}
\bibitem{Meyer:2004gx}
H.~B.~Meyer,
%``Glueball regge trajectories,''
[arXiv:hep-lat/0508002 [hep-lat]].
%127 citations counted in INSPIRE as of 09 Oct 2023
%\cite{Athenodorou:2020ani}
\bibitem{Athenodorou:2020ani}
A.~Athenodorou and M.~Teper,
%``The glueball spectrum of SU(3) gauge theory in 3 + 1 dimensions,''
JHEP \textbf{11} (2020), 172
doi:10.1007/JHEP11(2020)172
[arXiv:2007.06422 [hep-lat]].
%74 citations counted in INSPIRE as of 09 Oct 2023
%\cite{Huber:2021yfy}
\bibitem{Huber:2021yfy}
M.~Q.~Huber, C.~S.~Fischer and H.~Sanchis-Alepuz,
%``Higher spin glueballs from functional methods,''
Eur. Phys. J. C \textbf{81} (2021) no.12, 1083
[erratum: Eur. Phys. J. C \textbf{82} (2022), 38]
doi:10.1140/epjc/s10052-021-09864-5
[arXiv:2110.09180 [hep-ph]].
%27 citations counted in INSPIRE as of 09 Oct 2023
%\cite{Trotti:2022ukp}
%\cite{Gockeler:2005rv}
\bibitem{Gockeler:2005rv}
M.~Gockeler, R.~Horsley, A.~C.~Irving, D.~Pleiter, P.~E.~L.~Rakow, G.~Schierholz and H.~Stuben,
%``A Determination of the Lambda parameter from full lattice QCD,''
Phys. Rev. D \textbf{73} (2006), 014513
doi:10.1103/PhysRevD.73.014513
[arXiv:hep-ph/0502212 [hep-ph]].
%145 citations counted in INSPIRE as of 09 Oct 2023

%\cite{Giacosa:2021brl}
\bibitem{Giacosa:2021brl}
F.~Giacosa, A.~Pilloni and E.~Trotti,
%``Glueball\textendash{}glueball scattering and the glueballonium,''
Eur. Phys. J. C \textbf{82} (2022) no.5, 487
doi:10.1140/epjc/s10052-022-10403-z
[arXiv:2110.05582 [hep-ph]].
%6 citations counted in INSPIRE as of 09 Oct 2023





\end{thebibliography}
\end{document}